# Solving Maker-Breaker Games on 5-uniform hypergraphs is PSPACE-complete

Finn Koepke (finnkoepke@outlook.com)

**Abstract**

Let $(X, \mathcal{F})$ be a hypergraph. The Maker-Breaker game on $(X, \mathcal{F})$ is a combinatorial game between two players, Maker and Breaker. Beginning with Maker, the players take turns claiming vertices from $X$ that have not yet been claimed. Maker wins if she manages to claim all vertices of some hyperedge $F \in \mathcal{F}$. Breaker wins if he claims at least one vertex in every hyperedge.

M. L. Rahman and Thomas Watson proved in 2021 that, even when only Maker-Breaker games on 6-uniform hypergraphs are considered, the decision problem of determining which player has a winning strategy is PSPACE-complete. They also showed that the problem is NL-hard when considering hypergraphs of rank 5.

In this paper, we improve the latter result by showing that deciding who wins Maker-Breaker games on 5-uniform hypergraphs is still a PSPACE-complete problem. We achieve this by polynomial transformation from the problem of solving the generalized geography game on bipartite digraphs with vertex degrees 3 or less, which is known to be PSPACE-complete.

## 1 Introduction

Let $X$ be a finite vertex set and $\mathcal{F} \subseteq \mathcal{P}(X)$. Then, $(X, \mathcal{F})$ is a hypergraph and $\max_{F \in \mathcal{F}} |F|$ is its **rank**. Let $t \in \{\mathfrak{m}, \mathfrak{b}\}$. The **Maker-Breaker game** $(X, \mathcal{F}, t)$ is a combinatorial game between two players, Maker and Breaker. In the context of a Maker-Breaker game, we call the elements of $X$ **squares** and the elements of $\mathcal{F}$ **winning combinations**.

The game is played like this: The players take turns **claiming** a square from $X$ that has not yet been claimed. Maker goes first if $t = \mathfrak{m}$ and Breaker goes first if $t = \mathfrak{b}$. Maker wins if she manages to claim all squares of some winning combination $F \in \mathcal{F}$. The game ends with Breaker winning if all squares are claimed and Maker has not won, i.e. Breaker has claimed at least one square from every $F \in \mathcal{F}$.

If Maker has a winning strategy in the Maker-Breaker game $(X, \mathcal{F}, \mathfrak{m})$, we say that $(X, \mathcal{F})$ is **Maker's win**, otherwise it is **Breaker's win**. It is well-known folklore that if a hypergraph has multiple separate connected components, it is Maker's win if and only if one of the components is Maker's win. Therefore, we will only consider connected hypergraphs.

Let $(X, \mathcal{F})$ be a hypergraph, $X_M, X_B \subseteq X$ two disjoint sets of squares, and $t \in \{\mathfrak{m}, \mathfrak{b}\}$. The **position** $P = (X, \mathcal{F}, X_M, X_B, t)$ describes a gameplay state of a Maker-Breaker game on $(X, \mathcal{F})$ where Maker has claimed the squares in $X_M$ and Breaker has claimed the squares in $X_B$. If $t = \mathfrak{m}$, it is Maker's turn in $P$. Otherwise, it is Breaker's turn. In $P$, any winning combination $F \in \mathcal{F}$ with $X_B \cap F \neq \emptyset$ is no longer useful to Maker. We call such combinations **broken** in $P$.

One of the most useful qualities of Maker-Breaker games is that positions of larger Maker-Breaker games can be reduced to smaller Maker-Breaker games. Given $P = (X, \mathcal{F}, X_M, X_B, t)$, we can quickly construct a new Maker-Breaker game $(X_P, \mathcal{F}_P, t)$ that has the same gameplay as $P$ by letting $X_P := X \setminus (X_M \cup X_B)$ and $\mathcal{F}_P := \{F \setminus X_M \mid F \in \mathcal{F} \text{ not broken in } P\}$. We call $(X_P, \mathcal{F}_P, t)$ the **game that $P$ reduces to**.

Let $(X, \mathcal{F}, \mathfrak{b})$ be a Maker-Breaker game in which it is Breaker's turn. If $\mathcal{F}$ contains a winning combination $\{p\}$ of size one, we say that in $(X, \mathcal{F}, \mathfrak{b})$, **Maker is threatening mate in one**. This means that Breaker will lose unless he claims $p$. Similarly, if $\mathcal{F}$ contains two combinations $\{p, q\}$ and $\{p, r\}$ where $p$, $q$, and $r$ are pairwise distinct, we say that in $(X, \mathcal{F}, \mathfrak{b})$, **Maker is threatening mate in two**. Here, Breaker will lose unless he claims $p$, $q$ or $r$.

**Example.**

> Consider the Maker-Breaker version of Tic-Tac-Toe. Written formally, it looks like this:
>
> $(\{1, 2, ..., 8, 9\},\ \{\{1,2,3\}, \{4,5,6\}, \{7,8,9\}, \{1,4,7\}, \{2,5,8\}, \{3,6,9\}, \{1,5,9\}, \{3,5,7\}\}, \mathfrak{m})$
>
> The position after Maker claims 1, Breaker responds with 5, and Maker claims 9 reduces to the Maker-Breaker game $(\{1, 2, 4, 6, 7, 8\}, \{\{2, 3\}, \{7, 8\}, \{4, 7\}, \{3, 6\}\}, \mathfrak{b})$.
>
> | M | 2 | 3 |
> |---|---|---|
> | 4 | B | 6 |
> | 7 | 8 | M |
>
> In that game, Maker is threatening mate in two in two places at the same time: $2, 3, 6$ and $4, 7, 8$. Since these two sets of squares are disjoint, Breaker cannot stop both threats.

Let $(X, \mathcal{F})$ be a hypergraph. A **pairing** $C$ on $X$ is a collection of pairwise disjoint two-element subsets of $X$. We call the elements of $C$ **pairs**. We define the squares **covered by** $C$ as $\bigcup_{c \in C} c$. For each $F \in \mathcal{F}$, if there exists a $c \in C$ with $c \subseteq F$, we say that $C$ **blocks** $F$.

Given a pairing $C$, we can construct a **pairing strategy** for Breaker in the Maker-Breaker game $(X, \mathcal{F}, t)$. It goes as follows: Breaker responds to Maker claiming a square $p$ in one of two ways: If a pair $\{p, q\} \in C$ exists and $q$ is unclaimed, claim $q$. If that is not the case, he can claim any unclaimed square. If $t = \mathfrak{b}$, Breaker's first move can also be chosen arbitrarily.

By following this strategy, Breaker ensures that Maker cannot ever claim both squares of any pair $c \in C$. This also means that if $C$ blocks a winning combination $F$, Maker cannot win by claiming all squares in $F$. If $C$ blocks every winning combination in $\mathcal{F}$, its pairing strategy is a winning strategy for Breaker and we call $C$ a **complete pairing** of $(X, \mathcal{F})$. Any Maker-Breaker game played on a hypergraph that admits a complete pairing is Breaker's win (see Lemma 6 in [1]).

**Example.**

> Let $(X_n, \mathcal{F}_n) := (\{1, 2, ..., n-1, n\}, \{\{1, 2, 3\}, \{2, 3, 4\}, ..., \{n-2, n-1, n\}\})$ be a hypergraph.
>
> Then, $C_n := \{\{i, i+1\} \mid i \in X_n \setminus \{n\}, i \text{ odd}\}$ is a complete pairing of $(X_n, \mathcal{F}_n)$ because every winning combination $F \in \mathcal{F}_n$ is of the form $\{k-1, k, k+1\}$ for some $k \in X_n \setminus \{1, n\}$, and either $\{k-1, k\}$ or $\{k, k+1\}$ is contained in $C_n$.
>
> As a result, $(X_n, \mathcal{F}_n, \mathfrak{m})$ is Breaker's win.

## 2 Generalized Geography

The game *Geography* is a word game in which two players take turns naming geographical places. The starting word is fixed. For each place named, its first letter must be the same as the last letter of the previous word. The players may not repeat words. If a player cannot think of a valid word, they lose the game.

**Example.**

> Alice and Bob agree to play Geography only with the names of countries. They choose the starting word "Luxembourg". Alice goes first and has to say "Luxembourg", to which Bob answers "Germany". Alice must now say "Yemen" as it is the only country starting with a "y". Bob replies with "Norway" and Alice loses the game as she has no valid moves.

Geography can be generalized to a combinatorial game given by a weakly connected digraph $G = (V, A)$ and a starting vertex $s \in V$. The set of vertices corresponds to the set of allowed words, and each edge $(v, w)$ signifies that $w$ starts with the same letter $v$ ends in. An instance $(G, s)$ of **Generalized Geography** is therefore played as follows:

Alice begins by **marking** the designated starting vertex $s$. Then, starting with Bob, the two players alternate taking turns marking a previously unmarked vertex. It must be one that has an incoming edge from the previously marked vertex. The game ends when a player cannot make a legal move; that player loses the game.

**Notation.**

> Given a digraph $(V, A)$ and a vertex $v \in V$, we let $\delta^+(v)$ be the set of outgoing edges of $v$ and $\delta^-(v)$ be the set of incoming edges of $v$. Also, we let $\delta(v) := \delta^-(v) \cup \delta^+(v)$ be the set of all edges incident to $v$.

**Lemma 1.**

> The problem of deciding who wins an instance $(G, s)$ of Generalized Geography is PSPACE-complete even if we only consider the case where:
> 1. $G$ is planar and bipartite.
> 2. Each vertex $v \in V(G)$ fulfills $|\delta(v)| \leq 3$.
> 3. Each vertex $v \in V(G) \setminus \{s\}$ fulfills $|\delta^+(v)| \in \{1, 2\}$ and $|\delta^-(v)| \in \{1, 2\}$.
> 4. For the starting vertex $s$, we have $|\delta^+(s)| \in \{1, 2\}$ and $|\delta^-(s)| = 0$.

This lemma was proven in [2] by reduction from the true quantified boolean formula decision problem (TQBF). Even though points 3 and 4 were not explicitly stated, it is easy to verify that the constructed digraph (see Figure 1 in [2]) always has those properties. All these properties, besides planarity, will be very useful when constructing our Maker-Breaker game later.

**Corollary 2.**

> The problem from Lemma 1 remains PSPACE-complete even if we change point 4 such that it requires $s$ to have out-degree exactly 1.

**Proof.** Let $(G, s)$ be an instance of Generalized Geography that fulfills the properties in Lemma 1, but where $s$ has out-degree 2. Let $(s, v)$ and $(s, w)$ be the two outgoing edges of $s$. We can add two new vertices $x_1$, $x_2$ and replace the edges $(s, v)$ and $(s, w)$ with the edges $(s, x_1)$, $(x_1, x_2)$, $(x_2, v)$ and $(x_2, w)$. We call this new graph $G^+$.

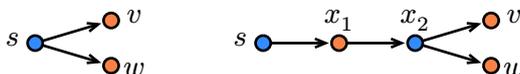

Doing so does not break planarity or 2-colorability. At the start of the game $(G^+, s)$, marking $s$, $x_1$, and $x_2$ is forced. The game state after these three vertices are marked in $(G^+, s)$ is completely identical to the state of the game $(G, s)$ after $s$ is marked. □

We call an instance $(G, s)$ of Generalized Geography **convertible** if it fulfills the requirements from Lemma 1 and Corollary 2. Given one such instance, a nice consequence of the 2-colorability of $G$ is that when playing Generalized Geography, the color of the most recently marked vertex $v$ indicates whose turn it currently is.

Because $G$ is weakly connected, choosing one of two colors for a single vertex yields a unique 2-coloring of $G$. Let $V_A \sqcup V_B = V(G)$ be the unique 2-coloring of $G$ with $s \in V_B$. As every edge runs between vertices of different colors, the color of the most recently marked vertex changes every turn. This means that if the most recently marked vertex $v$ has the same color as $s$, i.e. $v \in V_B$, it is Bob's turn. If $v \in V_A$, it is Alice's turn. Also, if a vertex in $V_A$ is marked, the player marking that vertex is Bob, and if a vertex in $V_B$ is marked, it must be marked by Alice.

## 3   Constructing the associated Maker-Breaker game

Let $G = (V, A)$ and $s \in V$ such that $(G, s)$ is a convertible instance of Generalized Geography. Again, let $V_A \sqcup V_B = V$ be the bipartition of $V$ such that $s \in V_B$. We can now partition the vertices of $G$ into six classes:

- Vertices in $V_A$ with in-degree 1 and out-degree 2 are in the class $M_{1,2}$.
- Vertices in $V_A$ with in-degree 2 and out-degree 1 are in the class $M_{2,1}$.
- Vertices in $V_B$ with in-degree 1 and out-degree 2 are in the class $B_{1,2}$.
- Vertices in $V_B$ with in-degree 2 and out-degree 1 are in the class $B_{2,1}$.
- Vertices with in-degree 1 and out-degree 1 are in the class $N_{1,1}$.
- The vertex $s \in V_B$ with in-degree 0 and out-degree 1 is in its own class $B_{0,1}$.

From $(G, s)$, we construct a Maker-Breaker game by first creating a separate hypergraph $H(v) = (X(v), \mathcal{F}(v))$ for each vertex $v \in V$, Then, for each edge $e = (u, w) \in A$, we identify a pair of squares in $X(u)$ with a pair of squares in $X(w)$.

More specifically, for each $v \in V$, let $H(v) = (X(v), \mathcal{F}(v))$ be a copy of the hypergraph in Table 1, column 3 of the row according to the class of $v$. Within this hypergraph, for each edge $e \in \delta(v)$, two squares of $X(v)$ are named $p_e$ and $q_e$, respectively. These squares are called **input squares** of $H(v)$ if $e \in \delta^-(v)$ and they are called **output squares** of $H(v)$ if $e \in \delta^+(v)$. Input squares are drawn with a red border in Table 1 and output squares are drawn with a green border. All other squares in $X(v)$ are called **interior squares** of $H(v)$ and are drawn with a blue border.

Let $e = (u, w) \in A$ be an edge. Then, $H(u)$ contains $p_e$ and $q_e$ as output squares and $H(w)$ contains $p_e$ and $q_e$ as input squares. We understand this pair of squares to be shared between $H(u)$ and $H(w)$, connecting the two hypergraphs. We call $p_e$ and $q_e$ the **joint squares** of $e$. On the other hand, interior squares $x \in X(v)$ of a hypergraph $H(v)$ are not shared and are therefore not contained in any other $X(w), w \neq v$.

Let $X := \bigcup_{v \in V} X(v)$ and $\mathcal{F} := \bigcup_{v \in V} \mathcal{F}(v)$. Then, $(X, \mathcal{F}, \mathfrak{m})$ is the **associated Maker-Breaker game** to $(G, s)$.

In Table 1, we describe how each class of vertex is handled during the construction of the associated Maker-Breaker game and during regular play. In column 2, we name the edges in $\delta(v)$. In column 3, we present the hypergraph $H(v)$. In column 4, we describe the order in which squares are claimed during regular play.

| Class of $v$ | $\delta(v)$ in $G$ | The hypergraph $H(v) = (X(v), \mathcal{F}(v))$ | Regular play |
|---|---|---|---|
| $v \in M_{1,2}$ | 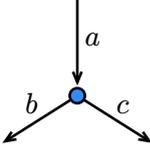 | 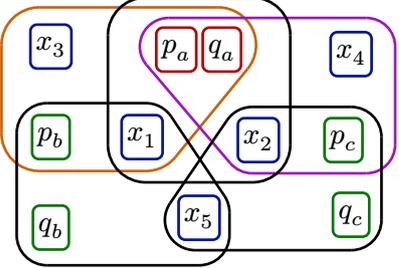 | If Maker chooses $b$: $x_1 \to x_2 \to p_b \to x_3$ $\to q_b \to x_5$ If Maker chooses $c$: $x_2 \to x_1 \to p_c \to x_4$ $\to q_c \to x_5$ |
| $v \in B_{1,2}$ | 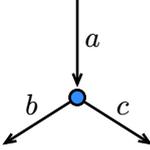 | 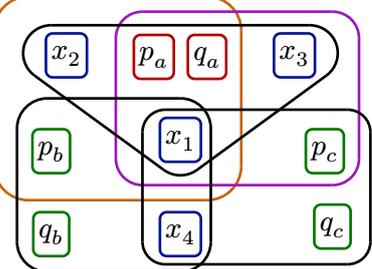 | If Breaker chooses $b$: $x_1 \to x_3 \to p_b \to x_2$ $\to q_b \to x_4$ If Breaker chooses $c$: $x_1 \to x_2 \to p_c \to x_3$ $\to q_c \to x_4$ |
| $v \in M_{2,1}$ | 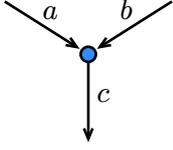 | 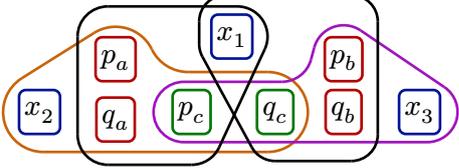 | If we enter through $a$: $p_c \to x_1 \to q_c \to x_2$ If we enter through $b$: $q_c \to x_1 \to p_c \to x_3$ |
| $v \in B_{2,1}$ | 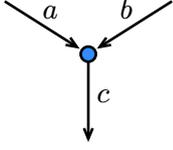 | 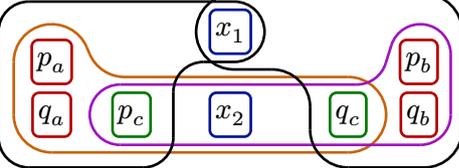 | If we enter through $a$: $p_c \to x_1 \to q_c \to x_2$ If we enter through $b$: $q_c \to x_1 \to p_c \to x_2$ |
| $v \in N_{1,1}$ | 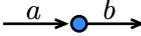 | 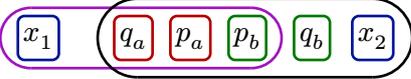 | $p_b \to x_1 \to q_b \to x_2$ |
| $v \in B_{0,1}$ | 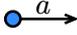 | 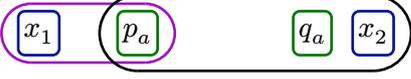 | $p_a \to x_1 \to q_a \to x_2$ |

Table 1: This table describes how we handle each class of vertex (see above). In column 3, squares outlined in red, green, and blue are input squares, output squares and interior squares of $H(v)$, respectively. The hyperedge colors have no special meaning and are just for visual clarity.

For each $v \in V$, we call the pairing $\{\{p_e, q_e\} \mid e \in \delta(v)\}$ on $H(v)$ its **joint pairing**. We observe that, if $v \neq s$, it is a complete pairing of the hypergraph $H(v)$. This also means that $\{\{p_e, q_e\} \mid e \in A\}$ is a pairing on $X$ and it blocks all winning combinations in $\mathcal{F} \setminus \mathcal{F}(s)$.

**Notation.**

Let $P = (X, \mathcal{F}, X_M, X_B, t)$ be a position of the associated Maker-Breaker and $v \in V$. Then, we define $X_P(v) := X_P \cap X(v)$ and $\mathcal{F}_P(v) := \{F \setminus X_M \mid F \in \mathcal{F}(v) \text{ not broken in } P\}$. We can imagine the hypergraph $(X_P(v), \mathcal{F}_P(v))$ as $(X_P, \mathcal{F}_P)$ restricted to $H(v)$.

## 4 Regular Play

Let $(G, s)$ be a convertible instance of Generalized Geography, and let $(X, \mathcal{F}, \mathfrak{m})$ be its associated Maker-Breaker game. Let $V(G) = V_A \sqcup V_B$ be the unique bipartition of $V(G)$ such that $s \in V_B$.

For each vertex class, column 4 of Table 1 shows its **regular play sequences**. For vertices in $M_{1,2}$ and $B_{1,2}$, there are two sequences because a choice is given to Maker or Breaker, respectively. For vertices in $M_{2,1}$ and $B_{2,1}$ there are two sequences because there are two possible incoming edges. Otherwise, there is only one sequence.

We can now formally describe **regular play** in $(X, \mathcal{F}, \mathfrak{m})$. The game begins with $s$ being what we call the **active vertex**. As long as there is an active vertex $v$, its class determines what happens next.

$$\text{Case 1: } v \in M_{1,2} \cup B_{1,2}$$

If $v \in M_{1,2}$, Maker chooses one of the two lines of play given in Table 1, row 1, column 4. If $v \in B_{1,2}$, Breaker chooses one of the two lines of play given in Table 1, row 2, column 4. Maker and Breaker play according to that line of play. Once these moves have been made, Maker will have claimed both joint squares of an outgoing edge $e$ of $v$. Which edge that is, depends on the chosen line of play. The next active vertex is $w$, where $e = (v, w)$.

$$\text{Case 2: } v \in N_{1,1} \cup B_{0,1}$$

Maker and Breaker follow the line of play in column 4 of the corresponding row in Table 1. Maker claims both joint squares of the edge $(v, w)$ leaving $v$. The next active vertex is $w$.

$$\text{Case 3: } v \in M_{2,1} \cup B_{2,1}$$

Here, $v$ has two incoming edges, $a$ and $b$. Let $e \in \{a, b\}$ be the edge that connects the previously active vertex to $v$, and let $c = (v, w)$ be the outgoing edge of $v$.

- If $v$ is active for the first time, depending on whether $e = a$ or $e = b$, Maker and Breaker follow the first or second line of play in Table 1, column 4 of $v$'s class. At its end, Maker will have claimed both joint squares of $c$. The next active vertex is $w$.

- If $v$ was active once already, the class of $v$ determines who will end up winning the game. The moment $v$ becomes active, it is Maker's turn and the squares $p_e$, $q_e$, $p_c$, and $q_c$ are already claimed by Maker. There is a winning combination $F \in \mathcal{F}(v)$ that contains all these squares plus an interior square $x$. If $v \in M_{2,1}$, $x$ is unclaimed. In that case, Maker claims $x$ and wins immediately. If $v \in B_{2,1}$, $x$ is already claimed by Breaker. In this case, there ceases to be an active vertex.

Once there is no active vertex anymore, regular play entails Maker claiming arbitrary squares and Breaker following the winning strategy provided to him via Lemma 3.

**Lemma 3.**

If Maker and Breaker follow regular play in $(X, \mathcal{F}, \mathfrak{m})$, and there ceases to be an active vertex because some $v \in B_{2,1}$ became active for a second time, that position is Breaker's win.

To prove this Lemma, we need to first establish some invariants that hold while there is an active vertex during regular play.

**Invariants of regular play.**

> Let $P$ be a position that occurs during regular play in $(X, \mathcal{F}, \mathfrak{m})$ while there is an active vertex $v$.
>
> 1. If it is Maker's turn in $P$, she will claim an interior square or an output square of $H(v)$. If it is Breaker's turn, he will claim an interior square of $H(v)$.
>
> 2. For each vertex $w$ that was never active, and every square $p \in X(w)$, one of these holds:
>    - The square $p$ is unclaimed in $P$.
>    - We have $(v, w) \in A$, $p$ is a joint square of that edge, and it is claimed by Maker.
>
> 3. For each vertex $u \neq v$ that was active previously, if $u \notin M_{2,1}$, we have $\mathcal{F}_P(u) = \emptyset$. In other words, as a result of regular play, every winning combination $F \in \mathcal{F}(u)$ is broken in $P$.
>
> 4. For each vertex $u \neq v$ that was active previously, if $u \in M_{2,1}$, $u$ has two incoming edges. Let $e'$ be the one that does *not* connect $u$ to the vertex that was active immediately before it. Then, we have $(X_P(u), \mathcal{F}_P(u)) = (\{p_{e'}, q_{e'}, x\}, \{\{p_{e'}, q_{e'}, x\}\})$.

**Proof of the invariants.** Invariant 1 can be verified by simply checking each of the regular play sequences in Table 1, column 4. Invariant 2 is a natural consequence of invariant 1.

The fact that invariant 3 holds for $u = s$ can be seen simply by checking Table 1. For $u \neq s$, let $u'$ be the vertex that was active immediately before $u$. Invariant 2 now tells us that when $u$ became active, the joint squares of $(u', u)$ were the only claimed squares in $X(u)$. Using this knowledge, we can use the table to confirm that the invariant 3 holds after each regular play sequence in $H(u)$ with $u \in M_{1,2} \cup B_{1,2} \cup B_{2,1} \cup N_{1,1}$.

Invariant 4 can be proven similarly: While $v \in M_{2,1}$ is active for the first time, regular play leaves the two input squares $p_{e'}, q_{e'}$ and one interior square $x$ of $H(v)$ unclaimed. There is a winning combination consisting of those three squares and the two output squares. The position after Maker claims the output squares reduces to $(\{p_{e'}, q_{e'}, x\}, \{\{p_{e'}, q_{e'}, x\}\})$. □

**Proof of Lemma 3.** Let $P$ be a position that occurs after $v \in B_{2,1}$ becomes active for a second time during regular play. We want to show that $C = \{\{p_e, q_e\} \mid e \in A \text{ with } \{p_e, q_e\} \subseteq X_P\}$ is a complete pairing of $(X_P, \mathcal{F}_P)$. It is obviously a pairing.

To show the pairing is complete, consider an arbitrary $F' \in \mathcal{F}_P$, and let $u \in V$ be a vertex such that $F' \in \mathcal{F}_P(u)$. If $u$ was active at some point, $u \in M_{2,1}$ must hold because otherwise, $\mathcal{F}_P(u) = \emptyset$ as per invariant 3. But then, invariant 4 gives us that $F'$ must have the form $\{p_e, q_e, x\}$ for some edge $e$, so $C$ blocks $F'$.

If $u$ was never active (which implies $u \neq s$), we know that $(v, u) \notin A$. Otherwise, $u$ would have become active immediately after $v$ was active for the first time. Paired with invariant 2, this means that all squares in $X(u)$ are unclaimed in $P$. Therefore, $C$ contains the joint pairing of $H(u)$. Since $u \neq s$, we know $C$ blocks $F'$. □

**Theorem 4.**

> If both players play perfectly while having to follow regular play in $(X, \mathcal{F}, \mathfrak{m})$, the outcome of that game is a victory for Maker if and only if $(G, s)$ is Alice's win.

**Proof.** First, we change the rules of Generalized Geography slightly. Marking a vertex more than once is no longer illegal. However, if a player marks an already marked vertex, that player loses the game. This change does not impact whether a given instance of Generalized Geography is Alice's win or Bob's win; we merely replaced a game loss from having no available legal moves with a game loss from having to mark an already marked vertex. We introduce these **revised rules** to more closely align the gameplay of $(G, s)$ with regular play in $(X, \mathcal{F}, \mathfrak{m})$.

We want to show that while there is an active vertex, regular play in $(X, \mathcal{F}, \mathfrak{m})$ is essentially identical to the gameplay in $(G, s)$ under revised rules. Alice is Maker, and Bob is Breaker. A vertex being most recently marked in $(G, s)$ is equivalent to it being active in regular play. The games start out in the same way: $s$ must be marked first in $(G, s)$, and it is the first active vertex during regular play.

In $(G, s)$, under revised rules, a player gets to make a choice if and only if it is their turn and the most recently marked vertex has more than one outgoing edge. During regular play of $(X, \mathcal{F}, \mathfrak{m})$, as long as there is an active vertex $v$, a player gets to make a choice if $v \in M_{1,2} \cup B_{1,2}$. These are exactly the vertices with more than one outgoing edge. The players who make the choices are identical as well: Maker or Alice gets to choose if $v \in V_A$ and Breaker or Bob gets to choose if $v \in V_B$.

In regular play, the winner is determined when a vertex $v$ becomes active for a second time. If $v \in V_A$, Maker wins by completing a winning combination, and if $v \in V_B$, Breaker has a winning pairing strategy as per Lemma 3. In $(G, s)$, under revised rules, the player who marks $v$ for a second time loses the game. We know from the 2-colorability of $G$ and from $s \in V_B$ that vertices in $V_A$ are always marked by Bob and vertices in $V_B$ are always marked by Alice. This means that the two games are decided under the same circumstances and with the same winners. If Alice wins $(G, s)$, Maker wins $(X, \mathcal{F}, \mathfrak{m})$, and vice versa.

All of this combined means that the gameplay in $(G, s)$ is essentially identical to the gameplay during regular play in its associated Maker-Breaker game. □

## 5 Irregular Play

In this section, we show that when playing the Maker-Breaker game $(X, \mathcal{F}, \mathfrak{m})$ associated with $(G, s)$, as long as there is an active vertex $v$, it is not beneficial for either player to violate the constraints set by regular play. Proving this is much easier for Breaker than for Maker.

**Lemma 5.**
> Let $v \in V(G)$ and let $P$ be a position that occurs during regular play of $(X, \mathcal{F}, \mathfrak{m})$ while $v$ is the active vertex and it is Breaker's turn. Then, if Breaker deviates from regular play, the resulting position is Maker's win.

**Proof.** By examining the regular play sequences in Table 1, we notice that unless regular play in $P$ specifically gives Breaker a choice, in $(X_P(v), \mathcal{F}_P(v))$, Maker is always threatening mate in one. In each of these cases, the move that prevents Maker from winning next turn is also the move that Breaker would make under regular play. This means that deviating from regular play allows Maker to win immediately.

In the case where Breaker has a choice, we have $v \in B_{1,2}$ and Maker has made one move since $v$ became active. Here, $\mathcal{F}_P(v)$ contains the winning combinations $\{p_b, x_2\}$, $\{x_2, x_3\}$ and $\{x_3, p_c\}$. This means that Maker is threatening mate in two in two places. The only way to stop both threats is to claim either $x_2$ or $x_3$. These are exactly the two choices Breaker has under regular play, so not following regular play causes the game to be Maker's win. □

To demonstrate that Maker also loses if she decides to deviate from regular play, we will prove that if she does so, Breaker always has a reply creating a position that admits a complete pairing. We will construct this complete pairing out of smaller ones.

Let $P = (X, \mathcal{F}, X_M, X_B, t)$ be a position and $v \in V$. Then, a pairing $C(v)$ on $(X_P(v), \mathcal{F}_P(v))$ is called a **puzzle piece pairing of** $(X_P(v), \mathcal{F}_P(v))$ if it has these traits:

1. It is a complete pairing of $(X_P(v), \mathcal{F}_P(v))$.

2. For each edge $e \in \delta(v)$, if $p_e \in X_P(v)$ and $q_e \in X_P(v)$, then $\{p_e, q_e\} \in C(v)$.

3. For each edge $e \in \delta^+(v)$, if $p_e \notin X_P(v)$ or $q_e \notin X_P(v)$, then $C(v)$ covers neither $p_e$ nor $q_e$.

**Lemma 6.**

> Let $v \in V \setminus \{s\}$ and $p \in X(v)$. Let $P = (X, \mathcal{F}, X_M, X_B, \mathfrak{m})$ such that $X_M \cap X(v) = \{p\}$ and $X_B \cap X(v) = \emptyset$. Then, there exists a puzzle piece pairing $C(v, p)$ of $(X_P(v), \mathcal{F}_P(v))$.

**Proof.** We call the squares in $X(v)$ by their names in Table 1, column 3.

$$\text{Case 1: } p \text{ is an interior square of } H(v).$$

If $p$ is an interior square, we know that the joint pairing $\{\{p_e, q_e\} \mid e \in \delta(v)\}$ is a complete pairing of $(X_P(v), \mathcal{F}_P(v))$, and it is also a puzzle piece pairing.

$$\text{Case 2: } p \text{ is an input square of } H(v), \text{ belonging to } e' \in \delta^-(v).$$

Let $q$ be the other joint square of $e'$. We know that $C^* := \{\{p_e, q_e\} \mid e \in \delta(v) \setminus \{e'\}\}$ should be a subset of $C(v, p)$ to fulfill trait 2 of puzzle piece pairings. If $v \in B_{1,2} \cup B_{2,1} \cup M_{2,1} \cup N_{1,1}$, we choose $C(v, p) = C^* \cup \{q, x_1\}$. If $v \in M_{1,2}$, we choose $C(v, p) = C^* \cup \{q, x_1\} \cup \{x_2, x_4\}$.

$$\text{Case 3: } p \text{ is an output square of } H(v), \text{ belonging to } e' \in \delta^+(v).$$

Again, $C^* := \{\{p_e, q_e\} \mid e \in \delta(v) \setminus \{e'\}\}$ should be a subset of $C(v, p)$. If $v \in B_{2,1} \cup M_{2,1} \cup N_{1,1}$, then $C^*$ is already a complete pairing of $(X_P(v), \mathcal{F}_P(v))$. If $v \in B_{1,2}$, let $C(v, p) = C^* \cup \{x_1, x_4\}$. If $v \in M_{1,2}$, we let $C(v, p) = C^* \cup \{x_1, x_5\}$ if $e' = b$ and $C(v, p) = C^* \cup \{x_2, x_5\}$ if $e' = c$.

As we can see, in every case, there is a puzzle piece pairing of $(X_P(v), \mathcal{F}_P(v))$. $\square$

**Lemma 7.**

> Let $P = (X, \mathcal{F}, X_M, X_B, \mathfrak{m})$ be a position of $(X, \mathcal{F})$. If, for each $v \in V$, the hypergraph $(X_P(v), \mathcal{F}_P(v))$ admits a puzzle piece pairing $C(v)$, then $C := \bigcup_{v \in V} C(v)$ is a complete pairing of $(X_P, \mathcal{F}_P)$.

**Proof.** To show that $C$ is a pairing, we must demonstrate that no square in $X_P$ is in more than one pair of $C$. This is automatically true for all interior squares since they can only occur in one of the puzzle piece pairings. Let therefore $p \in X_P$ be a joint square of an edge $(v, w)$. If $p_e$ and $q_e$ are both unclaimed in $P$, we have $\{p_e, q_e\} \in C(v)$ and $\{p_e, q_e\} \in C(w)$. If only $p$ is unclaimed, then we know that $C(v)$ does not cover $p$. In either case, there is only a single pair in $C$ that contains $p$.

To show that $C$ is complete, let $F \in \mathcal{F}_P$, and let $u \in V$ be the vertex such that $F \in \mathcal{F}_P(u)$. Then, since puzzle piece pairings are complete, $C(u)$ blocks $F$, so $C$ does too. $\square$

**Lemma 8.**

Let $v \in V(G)$ and let $P = (X, \mathcal{F}, X_M, X_B, \mathfrak{m})$ be a position that occurs during regular play of $(X, \mathcal{F})$ while $v$ is the active vertex and it is Maker's turn. Then, if Maker deviates from regular play, the resulting position is Breaker's win.

**Proof.** Within this proof, we refer to the square(s) that Maker could have played if she followed regular play in $P$ as the **regular play square(s)**.

Let $(X_P, \mathcal{F}_P)$ be the game $P$ reduces to and let $p \in X_P$ be any square besides the regular play square(s). We want to show that there exists a reply $q \in X_P \setminus \{p\}$ such that the resulting position $P' := (X, \mathcal{F}, X_M \cup \{p\}, X_B \cup \{q\}, \mathfrak{m})$ admits a complete pairing $C$.

| Class of $v$ | Since $v$ became active, Maker has claimed... | | |
|---|---|---|---|
| | ...no squares. | ...one square. | ...two squares. |
| $v \in M_{1,2}$ | 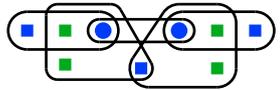 $\star$ | 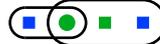 | 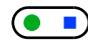 |
| $v \in B_{1,2}$ | 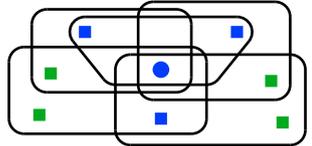 | 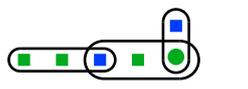 $\star\star$ | 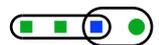 $\star\star$ |
| $v \in M_{2,1}$ | 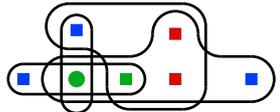 | 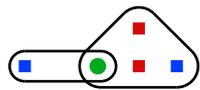 | |
| $v \in B_{2,1}$ | 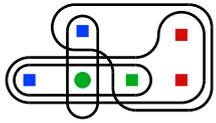 | 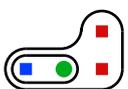 | |
| $v \in N_{1,1} \cup B_{0,1}$ | 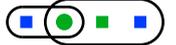 | 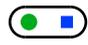 | |

Table 2: The possibilities for what $(X_P(v), \mathcal{F}_P(v))$ can look like. Input squares, output squares, and interior squares are red, green, and blue, respectively. The regular play squares are denoted by a circle shape instead of a square shape.

The choice of $q$ depends on which of these 10 forms the hypergraph $(X_P(v), \mathcal{F}_P(v))$ takes.

- If $(X_P(v), \mathcal{F}_P(v))$ takes any form besides those marked with a $\star$ or $\star\star$ in the bottom right corner of the cell, we always let $q$ be the regular play square.

- If $(X_P(v), \mathcal{F}_P(v))$ takes the form marked with $\star$, we similarly let $q$ be one of the two regular play squares. If $p$ shares a winning combination with one of them but not the other, we choose $q$ as the regular play square that does share a winning combination with $p$. Otherwise, both options for $q$ are equally viable.

- If $(X_P(v), \mathcal{F}_P(v))$ takes one of the forms marked with $\star\star$ (implying $v \in B_{1,2}$), let $e'$ be the outgoing edge where $\{p_{e'}, q_{e'}\}$ does not contain the regular play square. Here, it might be that case that Maker **tried to subvert Breaker's decision** by claiming a joint square of $e'$, i.e. $p \in \{p_{e'}, q_{e'}\}$. In that case, we let $q$ be the interior square of $H(v)$ that is unclaimed in $P$ and shares a winning combination with $p_{e'}$ and $q_{e'}$ (in Table 1, column 2 it is called $x_4$).

  If $p \notin \{p_{e'}, q_{e'}\}$, we simply let $q$ be the regular play square.

Let $(X_{P'}, \mathcal{F}_{P'})$ be the game that $P'$ reduces to. To show that a complete pairing $C$ of $(X_{P'}, \mathcal{F}_{P'})$ exists, we will use [Lemma 7](). This means that we try to find puzzle piece pairings of $(X_{P'}(w), \mathcal{F}_{P'}(w))$ for each $w \in V$. There are five categories that $w$ can fall into.

### Category 1: $w$ has been active before $v$ was, and $w \notin M_{2,1}$.

We know from [invariant 3 for regular play]() that $\mathcal{F}_P(w) = \emptyset$. It follows that $\mathcal{F}_{P'}(w) = \emptyset$, so $\{\{p_e, q_e\} \mid e \in \delta(v) \text{ with } p_e, q_e \in X_{P'}\}$ is a puzzle piece pairing of $(X_{P'}(w), \mathcal{F}_{P'}(w))$.

### Category 2: $w$ has been active before $v$ was, and $w \in M_{2,1}$.

We know from [invariant 4 for regular play]() that $(X_P(w), \mathcal{F}_P(w))$ takes the form $(\{p_e, q_e, x\}, \{\{p_e, q_e, x\}\})$ for some edge $e \in \delta^-(w)$ and some interior square $x$ of $H(w)$. If $q \in \{p_e, q_e\}$, then $\emptyset$ is a puzzle piece pairing of $(X_{P'}(w), \mathcal{F}_{P'}(w))$. If $p$ is one of these three squares and $q$ is not, then $\{X_P(w) \setminus \{p\}\}$ is a puzzle piece pairing of $(X_{P'}(w), \mathcal{F}_{P'}(w))$.

### Category 3: $w$ was never active and $X_P(w) = X(w)$.

If $p \in X_P(w)$, we know from [Lemma 6]() that a puzzle piece pairing $C(w, p)$ exists in $(X_{P'}(w), \mathcal{F}_{P'}(w))$. Otherwise, the joint pairing of $H(w)$ is a puzzle piece pairing of $(X_{P'}(w), \mathcal{F}_{P'}(w))$ after removing all pairs that contain $q$.

### Category 4: $w$ was never active, but $X_P(w) \neq X(w)$.

Then, we know from [invariant 2 for regular play]() that $e = (v, w) \in A$ and $X_P(w) = X(w) \setminus \{x\}$ for some $x \in \{p_e, q_e\}$. If Maker tried to subvert Breaker's decision, since $x$ is already claimed, we must be in the rightmost column of [Table 2](). We know that $p, q \notin X(w)$. Then, [Lemma 6]() tells us that a puzzle piece pairing $C(w, x)$ exists in $(X_{P'}(w), \mathcal{F}_{P'}(w))$.

If not, we chose $q$ to be the regular play square, which is the joint square of $e$ that is not $x$. If $p \notin X(w)$, we remove the pair $\{x, q\}$ from the joint pairing of $H(w)$. If $p \in X(w)$, we remove $\{x, q\}$ from $C(w, p)$ as provided by [Lemma 6](). Either way, we obtain a puzzle piece pairing of $(X_{P'}(w), \mathcal{F}_{P'}(w))$.

### Category 5: $w = v$.

If $(X_P(v), \mathcal{F}_P(v))$ takes a form besides the ones marked with $\star$ or $\star\star$, $\mathcal{F}_{P'}(v)$ is either empty or contains one winning combination. That winning combination contains three or four squares, depending on $p$. Here, we can find a puzzle piece pairing of $(X_{P'}(v), \mathcal{F}_{P'}(v))$ similarly to how we did in category 2.

If $(X_P(v), \mathcal{F}_P(v))$ takes the form marked with $\star$, let $q' \neq q$ be the regular play square we didn't choose for $q$. Then, $X_P(v)$ contains three squares that share a winning combination with $q'$ but not $q$, and therefore must be unclaimed: An interior square $x$ and two joint squares $p_e, q_e$ of some edge $e \in \delta^+(v)$. Therefore, we have $X_{P'}(v) \supseteq \{p_e, q_e, x, q'\}$. Looking at the remaining winning combinations, we see that $\{\{p_e, q_e\}, \{x, q'\}\}$ is a puzzle piece pairing of $(X_{P'}(v), \mathcal{F}_{P'}(v))$.

If $(X_P(v), \mathcal{F}_P(v))$ takes one of the two forms marked with $\star\star$ and Maker did not try to subvert Breaker's decision, we have $X_{P'}(v) \supseteq \{p_{e'}, q_{e'}\}$ and $\mathcal{F}_{P'}(v) = \{\{p_{e'}, q_{e'}\}\}$ for some $e' \in \delta^+(v)$. Here, $\{p_{e'}, q_{e'}\}$ is a puzzle piece pairing of that hypergraph.

If Maker did try to subvert Breaker's decision, $\mathcal{F}_{P'}(v)$ is either empty (if we are in the rightmost column of [Table 2]()) or, if we are in the middle column, it contains a winning combination of the form $\{x, p_e\}$, where $x$ is the interior square that is not $q$ and $e$ is the outgoing edge that contains the regular play square. Only in this case does $(X_{P'}(v), \mathcal{F}_{P'}(v))$ not admit a puzzle piece pairing, but it does admit the complete pairing $\{\{x, p_e\}\}$.

As we have seen, a puzzle piece pairing of $(X_{P'}(w), \mathcal{F}_{P'}(w))$ always exists for $w \neq v$, and it exists for $w = v$ in all cases but one. Unless we are in that case, the existence of a complete pairing of $(X_{P'}, \mathcal{F}_{P'})$ follows immediately from Lemma 7.

If we are in that case, we construct our complete pairing as follows: Let $e = (v, w')$ be the edge for which $p_e$ is the regular play square. For all $w \notin \{v, w'\}$, let $C(w)$ be the puzzle piece pairing given above. For the vertex $w'$, we know that it falls in category 1, 2, or 3, and $p \notin X(w')$. If it falls into category 1, let $C(w') = \emptyset$. If it falls into category 2, let $C(w') = \{\{q_e, x'\}\}$, where $x'$ is the one interior square in $X_{P'}(w')$. If it falls into category 3, let $C(w')$ be the puzzle piece pairing $C(w', p_e)$ obtained from Lemma 6. Finally, let $C(v) = \{\{p_e, x\}\}$, where $x$ is the one interior square of $X_{P'}(v)$. Then, $\bigcup_{w \in V} C(w)$ is a complete pairing of $(X_{P'}, \mathcal{F}_{P'})$. □

## 6 PSPACE-completeness

We can now put the pieces together and prove the result we have been working towards.

**Theorem 9.**

> Determining the winner of a Maker-Breaker game is a PSPACE-complete decision problem even if we only allow games played on hypergraphs of rank 5.

**Proof.** Let $(G, s)$ be a convertible instance of Generalized Geography, and $(X, \mathcal{F}, \mathfrak{m})$ its associated Maker-Breaker game. Combining Lemma 5 with Lemma 8 yields that one way for Maker and Breaker to play perfectly in $(X, \mathcal{F}, \mathfrak{m})$ is to follow regular play. As a result, $(X, \mathcal{F}, \mathfrak{m})$ is Maker's win if and only if it is also Maker's win when only perfect play is allowed. Adding Theorem 4, we obtain the result that $(G, s)$ is Alice's win if and only if $(X, \mathcal{F}, \mathfrak{m})$ is Maker's win.

Since $(X, \mathcal{F}, \mathfrak{m})$ can be constructed in linear time with respect to the size of $G$, we obtain a polynomial transformation from the problem of solving convertible instances of generalized geography to the problem of solving rank-5 Maker-Breaker games. Given Corollary 2, this means that the latter problem is PSPACE-hard.

Solving general Maker-Breaker games is a problem in PSPACE [3]. Hence, solving rank-5 Maker-Breaker games is a PSPACE-complete problem. □

**Corollary 10.**

> Determining the winner of a Maker-Breaker game is a PSPACE-complete decision problem even if we only allow games played on 5-uniform hypergraphs.

**Proof.** Let $(G, s)$ be a convertible instance of Generalized Geography, and $(X, \mathcal{F}, \mathfrak{m})$ its associated Maker-Breaker game. In the proof of Corollary 17 in [4], it was demonstrated how a hyperedge $F \in \mathcal{F}$ of size $n$ can be replaced by two hyperedges $F_1, F_2$ of size $n + 1$ without changing the winner of the Maker-Breaker game. We can use that method until all hyperedges have exactly size 5. The amount of edges we add scales only linearly with the size of $V(G)$, so the time complexity of constructing the associated Maker-Breaker game remains the same. □

# 7 Conclusion

By polynomial transformation from the problem of determining the winner of a special case of Generalized Geography, we could show that the problem of solving Maker-Breaker games on 5-uniform hypergraphs is also PSPACE-complete.

In [5], it was shown that the problem of solving Maker-Breaker games on hypergraphs of rank 3 or lower can be solved in polynomial time. This leaves us with an obvious open problem: How difficult is solving Maker-Breaker games on hypergraphs of rank 4?

## Acknowledgements

A version of this result began to take shape while I wrote my master's thesis. I am grateful to my supervisor Max Klimm for his great flexibility during that process. I also thank Florian Galliot for some fruitful exchanges. This paper was typeset using typst.